\documentclass[10pt]{article}    
\usepackage{amsmath,amssymb,latexsym,epsfig,subfigure,multirow,longtable,color}  
\usepackage{wrapfig}
\begin{document}
\pagenumbering{roman}

\clearpage \pagebreak \setcounter{page}{1}
\renewcommand{\thepage}{{\arabic{page}}}

\title{{\Large Editorial} \\
Mathematics at the eve of a historic transition in biology}

\author{Guo-Wei Wei$^{1,2,3}$ \footnote{ Address correspondences  to Guo-Wei Wei. E-mail:wei@math.msu.edu}\\
%\address{
$^1$ Department of Mathematics \\
Michigan State University, MI 48824, USA\\
$^2$  Department of Biochemistry and Molecular Biology\\
Michigan State University, MI 48824, USA \\
$^3$ Department of Electrical and Computer Engineering \\
Michigan State University, MI 48824, USA \\
}

\maketitle

%A century ago was an exciting time when physicists and mathematicians worked in tandem to establish quantum mechanism. Indeed, algebras, partial differential equations, group theory, and functional analysis underpin the foundation of quantum mechanism. Albert Einstein created geometrical theory of gravitation by using tensor fields defined on a Lorentzian manifold representing spacetime. Mathematics remains an indispensable tool in physics as evidenced by Nobel Prize in Physics 2016 for theoretical discoveries of topological phase transitions and topological phases of matter. Quantum filed theory and renormalization group still motivate much of current research in mathematics. 
\begin{wrapfigure}{r}{3in}
  \vspace{-4mm}
	 \includegraphics[keepaspectratio,width=3in]{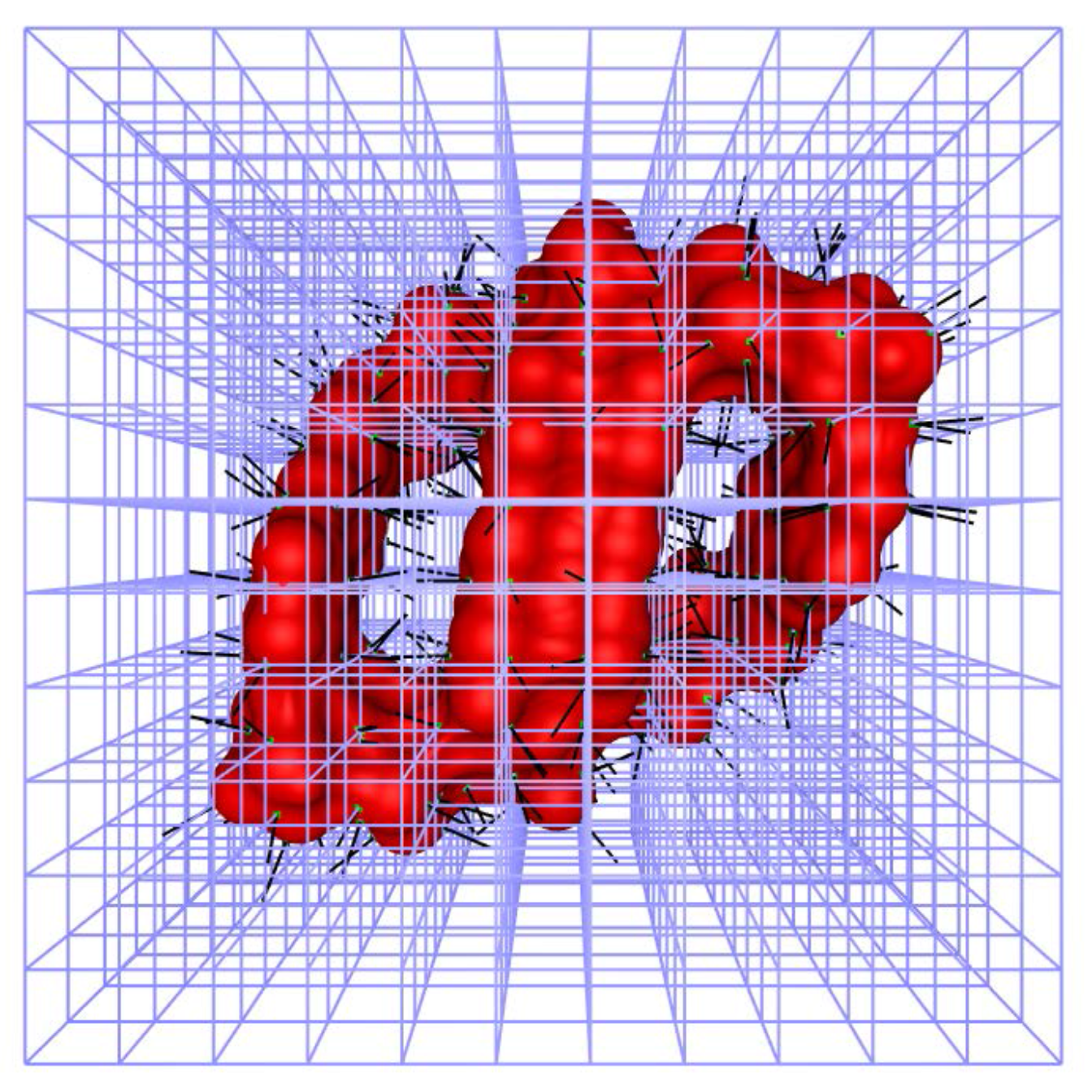}
    \vspace{-8mm}
	\caption{Geometric modeling of a protein surface in the Eulerian representation, which is crucial for the understanding of the protein structure and interactions, and bridges the gap between biomolecular structure data and mathematical models such as Poisson-Boltzmann equation or Poisson-Nernst-Planck equations.   
	Image credit: Rundong Zhao. 
	}
	\label{Fig}
	  \vspace{-4mm}
\end{wrapfigure}
Biology concerns the structure, function, development and evolution of living organisms. It underwent a dramatic transformation from macroscopic to microscopic (i.e.,``molecular'') in the 1960s and assumed an omics dimension around the dawn of the millennium. Understanding the rules of life is the major mission of biological sciences in the 21st century. The technological advances in the past few decades 
have fueled the exponential growth of biological data. For example, the Protein Data Bank (https://www.rcsb.org/pdb/home/home.do) has archived more than one hundred thirty thousand three-dimensional (3D) biomolecular structures and Genbank (https://www.ncbi.nlm.nih.gov/genbank/) has recorded more than 200 million sequences.
 The accumulation of various biological data in turn	has paved the way for biology to undertake another historic transition from being qualitative, phenomenological and descriptive to being quantitative, analytical and predictive.
%  Mathematics ought to be the driving force behind this historic transition as it did for quantum physics a century ago. 
Such a transition 
%The current challenges in understanding the rules of life offer 
provides both unprecedented  opportunities and grand challenges for mathematicians.  

\begin{wrapfigure}{r}{3in}
  \vspace{-4mm}
	 \includegraphics[keepaspectratio,width=3in]{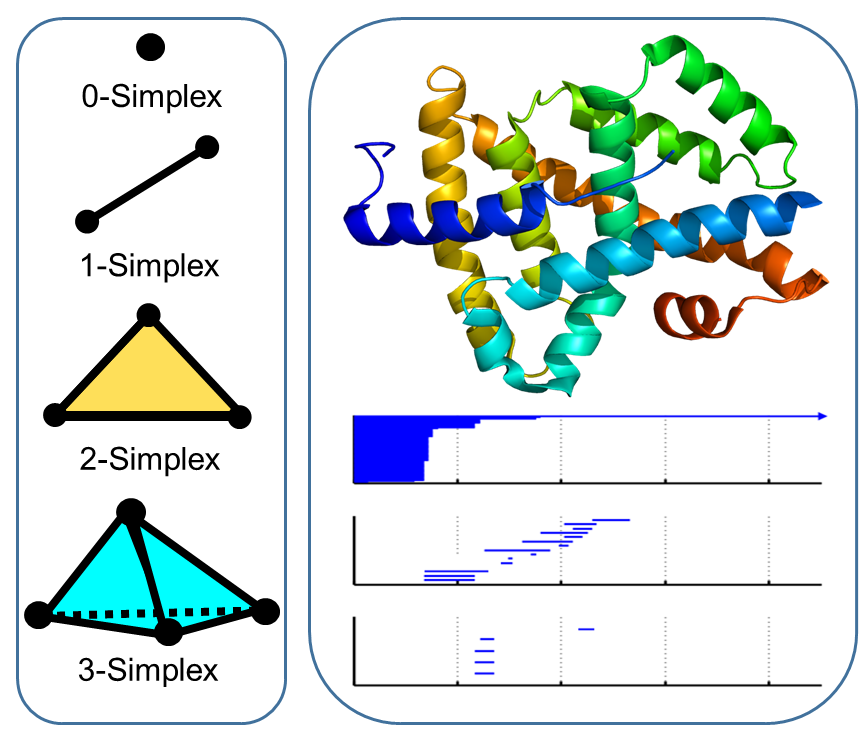}
    \vspace{-8mm}
	\caption{Basic simplexes (Left column) and protein persistent barcodes (Right column).
	Image credit: Zixuan Cang. 
	}
	\label{Fig2}
	   \vspace{-3mm}
	\end{wrapfigure}

One of major challenges in biology is the understanding of structure-function relationships in biomolecules,  such as proteins, DNA, RNA, and their interacting complexes. Such an understanding is the holy grail of biophysics and has a profound impact to biology, biotechnology, bioengineering and biomedicine.  
Mathematical %ly, geometric, topological and graph theory 
apparatuses, including simplicial geometry, differential geometry, differential topology, algebraic topology, geometric topology, knot theory, tiling theory, spectral graph theory and topological graph, are essential for %quantitatively describing biomolecular structures, modeling biological functions, and 
deciphering biomolecular structure-function relationships \cite{KLXia:2016,Twarock:2008,Jonoska:2009,RBrasher:2013}. 
In general, geometric modeling  is paramount  for the conceptualization of  biomolecules and their interactions, which is vital to the understanding of intricate  biomolecules. 
Geometric modeling also bridges the gap between biological data and mathematical models  involving topology, graph theory and partial differential equations (PDEs) \cite{ZYu:2008,ESES:2017} (see Figure \ref{Fig}).
 Topology dramatically simplifies biological complexity   and renders  insightful high level abstraction to large 
biological data \cite{Zomorodian:2005,sumners:1992,YaoY:2009, IKDarcy:2013, CHeitsch:2014} (see Figure \ref{Fig2}).  Graph theory is able to go beyond    topological connectivity and incorporates harmonic analysis and optimization theory to explore biomolecular  structure-function relationships. 

%The dynamics and transport of biomolecules in living organisms, such as membrane transport, signal transduction, transcription and translation   are essential to cellular functions and involve material and energy balances. Mathematical modeling provides an effective description of dynamics and transport in cells.  For example, multiscale  dynamical and transport equations describing the advection, diffusion and reaction of biomolecules can be incorporated with discrete atomic/molecular information, such as atomic charge, rigidity, mobility, et cetera, to describe  dynamics and transport in cells \cite{Eisenberg:2010,chen2016new,CaiW:2016}. 

 A striking feature of living organisms  is their multiscale nature and tremendous complexity. Subcellular organelles, molecular machines, and  
%, and multiprotein complexes 
dynamics and transport of biomolecules in living organisms, such as membrane transport, signal transduction, transcription and translation,   
are vital to cellular functions and cannot be simply described by atom-free or molecule-free phenomenological models. However, at the atomic scale, these systems  have intractable numbers of degrees of freedom. Multiscale modeling and analysis using quantum mechanics (QM), molecular mechanics (MM), and continuum mechanics (CM) offer an effective reduction in their dimensionality \cite{Eisenberg:2010,chen2016new, CaiW:2016}. 
PDEs (e.g., Schrodinger equation, Poisson-Boltzmann equation, elasticity equation etc.), Newton's equations of motion, variational analysis, homogenization, differential geometry, persistently stable manifolds, et cetera, underpin multiscale QM/MM/CM modeling of excessively large biological systems  \cite{Zhou:2008d,Wei:2009,geng2017two}. 
Differential geometry theory of surfaces gives rise to a natural separation between microscopic and macroscopic domains \cite{Wei:2009,ZhanChen:2012}. %Mathematical techniques for conservation law systems can be adopted to deal with the dynamics and transport of biomolecular systems. Stochastic analysis and uncertainty quantification can be utilized to reveal how individual biomolecular behavior is related to macroscopic measurements.   
Conservation law, %can be adopted to deal with the dynamics and transport of biomolecular systems. 
stochastic analysis and uncertainty quantification can be utilized to reveal how individual biomolecular behavior is related to experimental  measurements \cite{lei2015constructing}. 
A major challenge is how to maintain adequate descriptions of biological properties of interest defined by a given sets of experimental measurements, while dramatically reducing the dimensionality of the underlying biological systems.

It remains poorly understood how various macromolecular complexes interact and give rise to cellular functions and biological pathways, e.g.,  metabolic, genetic and signal transduction pathways \cite{komarova2005theoretical}. Differential equations, combinatorics, probability graph, random matrix, statistical models and algebraic geometry are the main workhorses for describing interactive biological networks, such as protein-protein interaction, gene regulation, and enzyme kinetics networks.  The systems biology approaches often involve mechanistic models, such as flux balance analysis and chemical kinetics, to reconstruct the dynamical systems from the quantitative properties of their elementary building blocks. Computational biophysics predict reaction flux, rate and equilibrium constants of biological networks.

In the post-omics era, the availability of high-throughput sequencing strategies have resulted in  genomics, proteomics and metabolomics. 
%enormous whole genomic data. 
Omics aims at the integrative studies of whole set of  biomolecular information  that translates into the structure, function, and evolution of living organisms.  One major challenge is how to predict phenomics from genomics aided by ChIA-PET and/or trait information.  Another challenge is the understanding of  genome evolution due to gene-gene and gene-environment interactions. Statistical methods, such as longitudinal study, causal analysis, statistical inference, fuzzy logic, boosting  and  regression, play a vital role in analyzing omic data sets. Machine learning is another powerful tool for revealing the genotype to phenotype mapping \cite{newton2001differential}.  
Mathematical, statistical, and machine learning  approaches are essential for the understanding and prediction of genetic data, such as gene sequencing, expression, regulation, evolution, mutation and biological pathways so as to result in  therapeutic benefit to patients \cite{yu2017personalized}.

Structural bioinformatics concerns the modeling, analysis and prediction of biomolecular structural properties, including protein folds, protein pKa, mutation induced free energy changes, binding affinities of protein-protein and protein-nucleic acid interactions, and the 3D structures of RNA and RNA-protein complexes \cite{CHeitsch:2014}. Mathematical techniques using geometry, topology and graph theory have a competitive edge in structural bioinformatics \cite{Demerdash:2009, ZXCang:2017a}. For example, persistent homology strikes a balance between biological detail and topological simplification to achieve appropriate abstractions to massive 3D structural data \cite{ZXCang:2017c}.

The importance of biotechnologies for 3D structure determination and for nucleotide sequencing to biological sciences cannot be over emphasized. High-throughput sequencing methods based on chemical synthesis, fluorescent labeling, capillary electrophoresis, and general automation have fundamentally changed molecular biology, evolutionary biology metagenomics, medicine, forensics and anthropology.  Macromolecular X-ray crystallography, nuclear magnetic resonance (NMR), cryo-electron microscopy (cryo-EM), electron paramagnetic resonance (EPR), multiangle light scattering, confocal laser-scanning microscopy, scanning capacitance microscopy, small angle scattering, ultra-fast laser spectroscopy, et cerate, determine 3D structures of macromolecules. Mathematics, such as harmonic analysis, approximation theory, Tikhonov-regularization,  inverse scattering theory, et cetera, plays an important role in advancing biotechnologies. For example, cryo-EM is one of the most promising techniques for the structure determination of excessively large biomolecules and relies on mathematical algorithms for image analysis and structure reconstruction \cite{singer2011three}.  Additionally, the accuracy of electrophoresis based sequencing can be improved through mathematical modeling of microfluidic and nanofluidic devices.

The development of accurate, efficient and robust computational algorithms, methods and schemes is a prerequisite for the implementation of  mathematical approaches to biological modeling, analysis and prediction. The importance of numerical methods in solving PDEs is  gradually appreciated by the biological community \cite{SZhao:2011,Geng2013:3,xied:2013}. Computational geometry is an important aspect in structural biology and biophysics \cite{ESES:2017}. Computational topology analyzes the intriguing topology of complex  biomolecules, such as topological invariants of proteins  and knot invariants of nucleosomes and chromosomes \cite{Schlick:1992trefoil}. The development of efficient graph theory algorithms is crucial for the description of biomolecular binding  \cite{DDNguyen:2017d}.  Advanced statistic methods underpin various bioinformatic predictions, including 3D protein folds from sequence information.   Deep learning algorithms promise the discovery of geometry-function relationships and topology-function relationships from massive  biological data.    
%have remarkably boosted  

Rational drug design is an imperative life science problem that ultimately tests out our understanding of biological systems. Designing efficient drugs for curing diseases is one of the most challenging tasks in biological sciences. 
It involves a complex procedure, including disease identification, target hypothesis (i.e., the activation or inhibition of drug targets), screening of potential drugs that can effectively bind to the target while having low affinity to off-targets, optimization of the structures of selected drug candidates, {\it in vitro} and {\it in vivo} preclinical tests, clinical trials to examine bioavailability and therapeutic potential, and the optimization of a drug's efficacy, toxicity, and pharmacokinetic properties. 
Mathematics   plays a crucial role in hot-spot prediction, drug pose analysis, binding affinity prediction,   structure optimization, toxicity analysis and pharmacokinetic simulation \cite{BaoWang:2016FFTB}. For example,  the integration of machine learning with multiscale weighted colored graphs  and multicomponent persistent homology  provided the best free energy ranking for Set 1 (Stage 2) in  D3R Grand Challenge 2, a world-wide competition in computer aided drug design (http://users.math.msu.edu/users/wei/D3RFreeEnergy.pdf).   It is expected that most new drugs in the next decade will be initiated by artificial intelligence.  

\end{document}